\documentclass[a4paper]{article}

\usepackage{INTERSPEECH2022}
\usepackage{subfigure}
\title{Cross-Scale Vector Quantization for Scalable Neural Speech Coding}
\name{Xue Jiang$^{1\ast}$, Xiulian Peng$^2$, Huaying Xue$^2$, Yuan Zhang$^1$, Yan Lu$^2$}
\address{
  $^1$Communication University of China, Beijing, China\\
  $^2$Microsoft Research Asia, Beijing, China}
\email{jiangxhoho@cuc.edu.cn,xipe@microsoft.com,
huxue@microsoft.com,yzhang@cuc.edu.cn,yanlu@microsoft.com}

\begin{document}

\maketitle
\renewcommand{\thefootnote}{\fnsymbol{footnote}}
\footnotetext[1]{This work was done when Xue Jiang was an intern at MSRA.} 

\begin{abstract}
Bitrate scalability is a desirable feature for audio coding in real-time communications. Existing neural audio codecs usually enforce a specific bitrate during training, so different models need to be trained for each target bitrate, which increases the memory footprint at the sender and the receiver side and transcoding is often needed to support multiple receivers. In this paper, 
we introduce a cross-scale scalable vector quantization scheme (CSVQ), in which multi-scale features are encoded progressively with stepwise feature fusion and refinement. In this way, a coarse-level signal is reconstructed if only a portion of the bitstream is received, and progressively improves the quality as more bits are available. The proposed CSVQ scheme can be flexibly applied to any neural audio coding network with a mirrored auto-encoder structure to achieve bitrate scalability. Subjective results show that the proposed scheme outperforms the classical residual VQ (RVQ) with scalability. Moreover, the proposed CSVQ at 3 kbps outperforms Opus at 9 kbps and Lyra at 3kbps and it could provide a graceful quality boost with bitrate increase.
\end{abstract}

\section{Introduction}
Audio coding typically employs a carefully-designed pipeline to remove the redundancy in the source signal and yield a compact bitstream. The goal of audio coding is to represent a audio signal with minimum number of bits while retaining its quality. Recently many deep learning-based methods have been proposed for audio coding and achieved very promising results. Some researchers leverage advances in speech synthesizing with generative models \cite{kleijn2018wavenet,klejsa2019high,skoglund2019improving,fejgin2020source, oord2016wavenet, kleijn2021generative,valin2019lpcnet, mehri2016samplernn}, such as WaveNet \cite{oord2016wavenet}, its variants WaveGRU in Lyra \cite{kleijn2021generative}, LPCNet \cite{valin2019lpcnet} and SampleRNN \cite{mehri2016samplernn}. They typically utilize a powerful generative decoder model conditioned on handcrafted acoustic features extracted from a speech signal. On the other hand, some researchers propose end-to-end neural networks based on the vector-quantized variational autoencoder framework (VQ-VAE \cite{oord2017neural}) where the encoder, codebook and the decoder are learned in a joint fashion \cite{kankanahalli2018end,garbacea2019low,cascaded,casebeer2021enhancing,zeghidour2021soundstream,TFNetCodec}. These methods encode the input signal into a discrete representation and then reconstruct the original signal from these latent sequences. These methods have demonstrated the ability of deep neural networks to produce high quality audio at a low bitrate. However, they usually enforce a specific bitrate during training and require retraining multiple models for different target bitrates. 

\begin{figure}[tb]
\centering
\begin{minipage}[b]{0.93\linewidth}
  \centering
  \centerline{\includegraphics[width=1.0\linewidth]{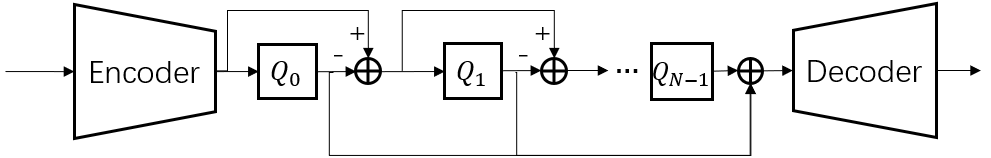}}
  \centerline{(a) Single-scale residual VQ \cite{vasuki2006review}}\medskip
\end{minipage}
\centering
\begin{minipage}[b]{0.93\linewidth}
  \centering
  \centerline{\includegraphics[width=1.0\linewidth]{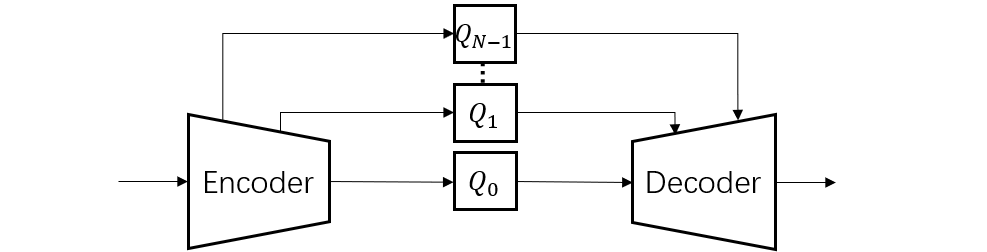}}
  \centerline{(b) Multi-scale VQ \cite{petermann2021harp}}\medskip
\end{minipage}
\centering
\begin{minipage}[b]{0.93\linewidth}
  \centering
  \centerline{\includegraphics[width=1.0\linewidth]{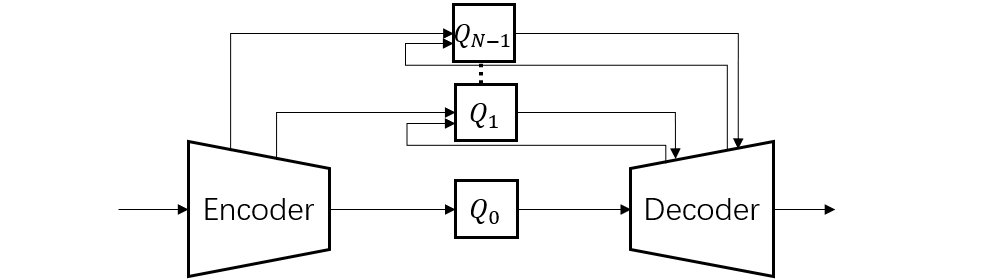}}
  \centerline{(c) Our cross-scale VQ }\medskip
\end{minipage}
\vspace{-0.4cm}
\caption{Scalable vector quantization.}
\vspace{-0.2cm}
\label{fig:SVQ}
\vspace{-0.5cm}
\end{figure}

Bitrate scalability is a desirable feature in coding, especially for streaming and real-time communications. Some researchers introduce residual vector quantization (RVQ \cite{vasuki2006review}) into audio coding to achieve bitrate scalability \cite{cascaded,zeghidour2021soundstream,zhen2020efficient,zhen2021scalable}. RVQ provides a convenient framework for controlling the bitrate, where each quantizer quantizes the residual from the previous stage and thus gradually improves the quality of the quantized vector as shown in Figure \ref{fig:SVQ}(a). However, the $N$ quantizers in RVQ all perform on a fixed-scale feature, typically the output of the encoder, ignoring the multi-scale information from different layers of the encoder. Some researchers \cite{petermann2021harp} replace the identity shortcuts in the original U-Net with additional autoencoders and deliver the multi-scale features in the compressed form to the decoder side, as shown in Figure \ref{fig:SVQ}(b). The bitstreams at multiple scales promote the information communication between encoder and decoder so that better layer-wise approximation to the encoder feature is achieved in the decoder which finally leads to better output quality. However, there is no explicit dependency between these bitstreams where the redundancy between them is not controlled. 

\begin{figure*}
\centering
\begin{minipage}[b]{0.6\linewidth}
    \subfigure[Network structure]{\includegraphics[width=1\linewidth]{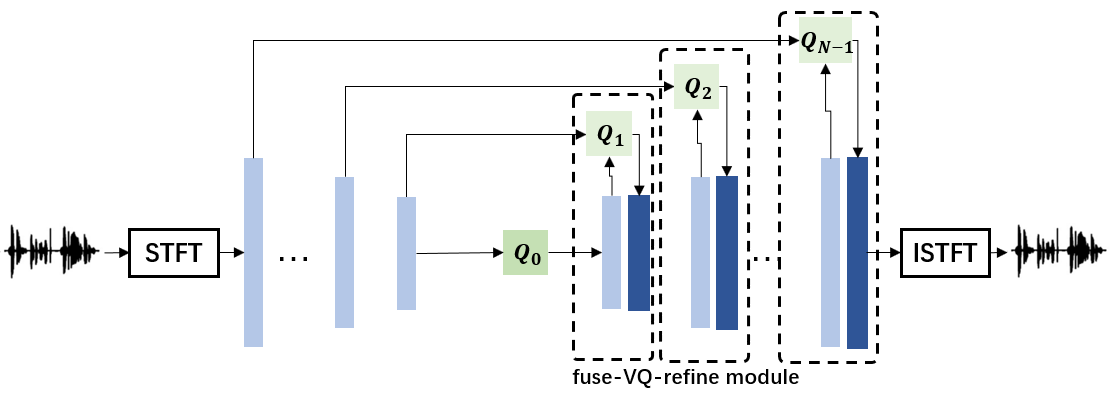}}
    \end{minipage}
\begin{minipage}[b]{0.37\linewidth}
    \subfigure[Fuse-VQ-refine module: CSVQ-merge]{\includegraphics[width=1\linewidth]{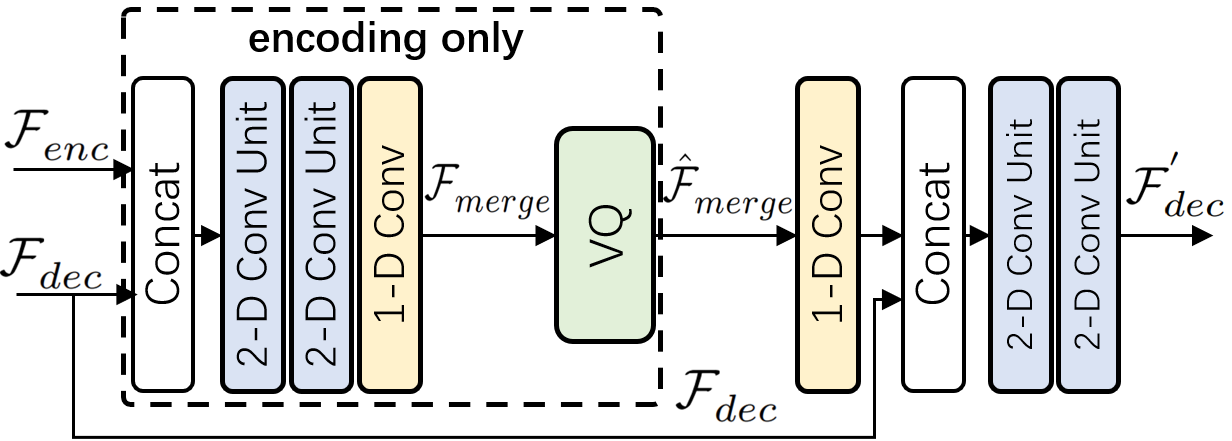}}
    \subfigure[Fuse-VQ-refine module: CSVQ-residual]{\includegraphics[width=1\linewidth]{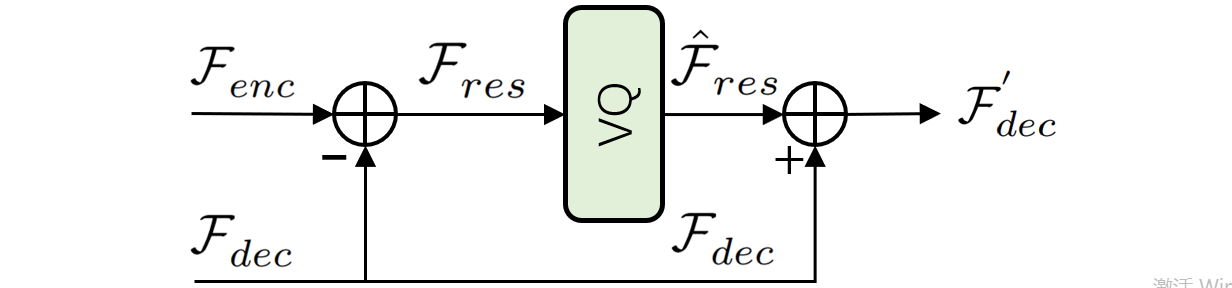}}
\end{minipage}
\vspace{-0.4cm}
\caption{The overall network structure. A convolutional encoder produces a compact latent representation of the input signal along with multi-scale features, which are quantized using a variable number of vector quantizers after fusion with reconstructed features from decoding. The fully convolutional decoder receives each quantized embedding, refines the reconstructed feature from previous bitstreams with it and finally reconstructs the whole signal.}
\vspace{-0.5cm}
\label{fig:model structure}
\end{figure*}
In this paper, we propose the cross-scale scalable vector quantization scheme (CSVQ) as shown in Figure \ref{fig:SVQ}(c). Different from that in Figure \ref{fig:SVQ}(b), the additional short cuts with bottleneck VQs encode the layer feature of the encoder conditioned on the decoder feature that is produced from previous bitstreams. This explicitly removes the redundancy between different bitstreams and helps to boost the rate-distortion performance. The additional information by the short cut is then merged with the decoder feature for refinement. In CSVQ, the base quantizer $Q_0$ produces the bitstream with more high-level information and other quantizers $Q_1, Q_2, ..., Q_{N-1}$ produce additional bitstreams containing more detailed high-frequency information. This fuse-VQ-refine module by CSVQ could be flexibly applied to any neural audio coding network with a mirrored autoencoder structure. 

Taking the low-latency neural speech codec TFNet \cite{TFNetCodec} as our backbone, we introduce the bitrate-scalable TFNet (S-TFNet) with the proposed CSVQ scheme, which can operate across variable bitrates from 3 kbps to 18 kbps with a single model. The S-TFNet is end-to-end trained in a single stage by randomly sampling a bitrate for each minibatch. Experimental results show that CSVQ outperforms the RVQ subjectively and the S-TFNet at 3kbps outperforms Opus \cite{valin2012definition} at 9kbps and Lyra \cite{kleijn2021generative} at 3kbps with a graceful rate-distortion trade-off. Moreover, the scalable S-TFNet optimized for multiple bitrates performs on par with that optimized for a fixed bitrate, showing the efficiency of the one-stage training algorithm.

\section{The Proposed Scheme}

\subsection{Overview}
Taking TFNet \cite{TFNetCodec} as the backbone, the proposed S-TFNet consists of an encoder, several vector quantizers $Q_0, Q_1, ..., Q_{N-1}$, and a decoder, as shown in Figure \ref{fig:model structure}(a). The encoder takes complex time-frequency (T-F) spectrum as input and produces a series of features at different scales, which are then quantized selectively according to the target bitrate after a fusion with each reconstructed feature from the decoder. The decoder receives each quantized embedding, refines the reconstructed feature from previous bitstreams with the new embedding and gradually reconstructs the signal. Causal convolutions are used for the whole network so that it could keep a low latency of 20ms when STFT uses a window of 20ms with a 5ms hop length. The following subsections will explain each part in detail.

\subsection{Encoder and Decoder}
\label{ssec:encoder and decoder}
We modify the TFNet structure a little bit to facilitate more bitrate scalability within a single model. The encoder consists of six causal 2-D convolutional layers, followed by a temporal convolution module (TCM) similar to that in \cite{pandey2019tcnn} and a GRU layer for long-term temporal correlation exploitation. It takes complex spectrum given by short-time Fourier transform (STFT) as the input, denoted by $X^{I}\in \mathbb{R}^{T\times F \times 2}$, where $T$ is the number of frames and $F$ is the number of frequency bins. The six 2-D convolutional layers successively reduce the size along the frequency dimension with a stride of 2 and finally all frequency information is folded into channels. No down-sampling is performed along the temporal dimension to preserve the temporal resolution. After the convolutional layers, the reshaped feature $X\in \mathbb{R}^{T\times 1 \times C}$ is fed into a TCM module followed by a GRU layer which learns a high-level feature by exploring long-range dependencies from the past frames.

The decoder consists of two large TCM modules and one GRU layer in an interleaved manner, followed by a series of 2-D transposed convolutional layers. The interleave TCM and GRU structure captures long-term dependencies from the past to help to reconstruct the orignal signal. The following six 2-D transposed convolutional layers are a mirror-image of convolutional layers at the encoder and each deconvolutional layer is followed by a refinement module in the CSVQ, performing a stepwise refinement. After the decoder, an inverse STFT is applied to reconstruct the waveform audio.

\begin{figure*}[tb]
\centering
\begin{minipage}{1.0\linewidth}
    \begin{minipage}[b]{0.33\linewidth}
      \centering
      \subfigcapskip=-3pt
      \subfigure[MUSHRA score (low bitrate)]{\includegraphics[width=5.75cm]{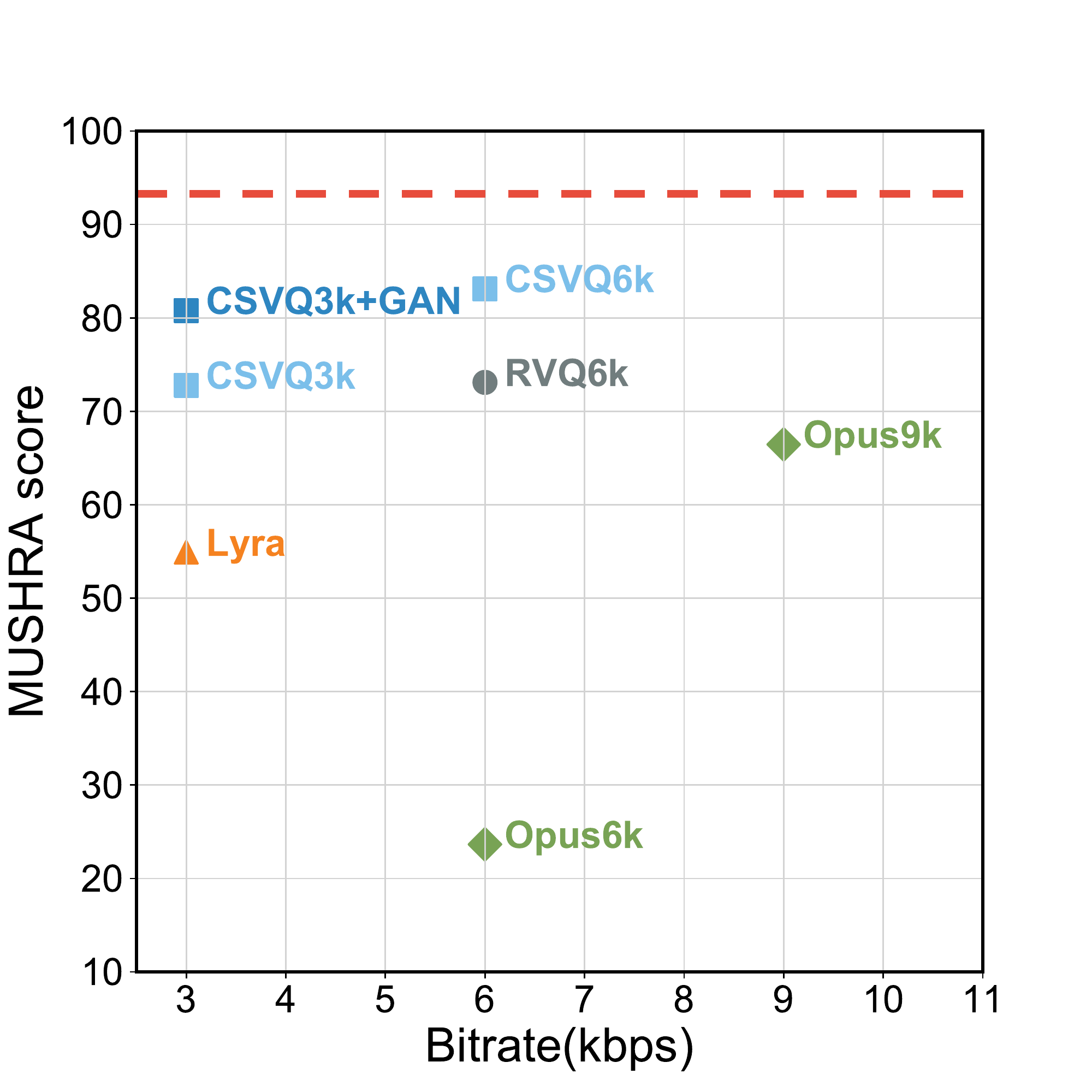}}
    \end{minipage}
    \begin{minipage}[b]{0.33\linewidth}
      \centering
      \subfigcapskip=-3pt
      \subfigure[MUSHRA score (high bitrate)]{\includegraphics[width=5.75cm]{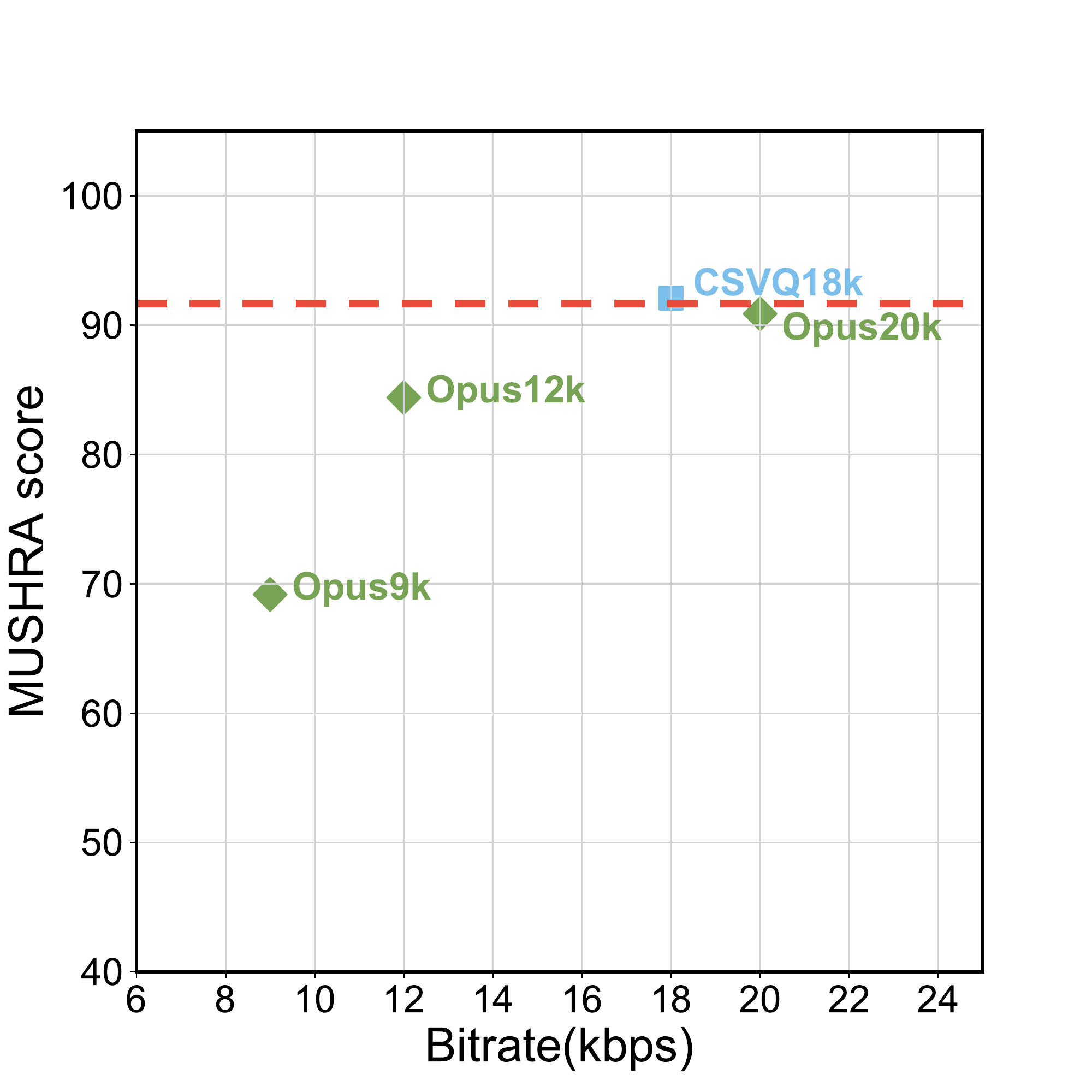}}
    \end{minipage}
    \begin{minipage}[b]{0.33\linewidth}
      \centering
      \subfigcapskip=-3pt
      \subfigure[PESQ vs. bitrate]{\includegraphics[width=5.75cm]{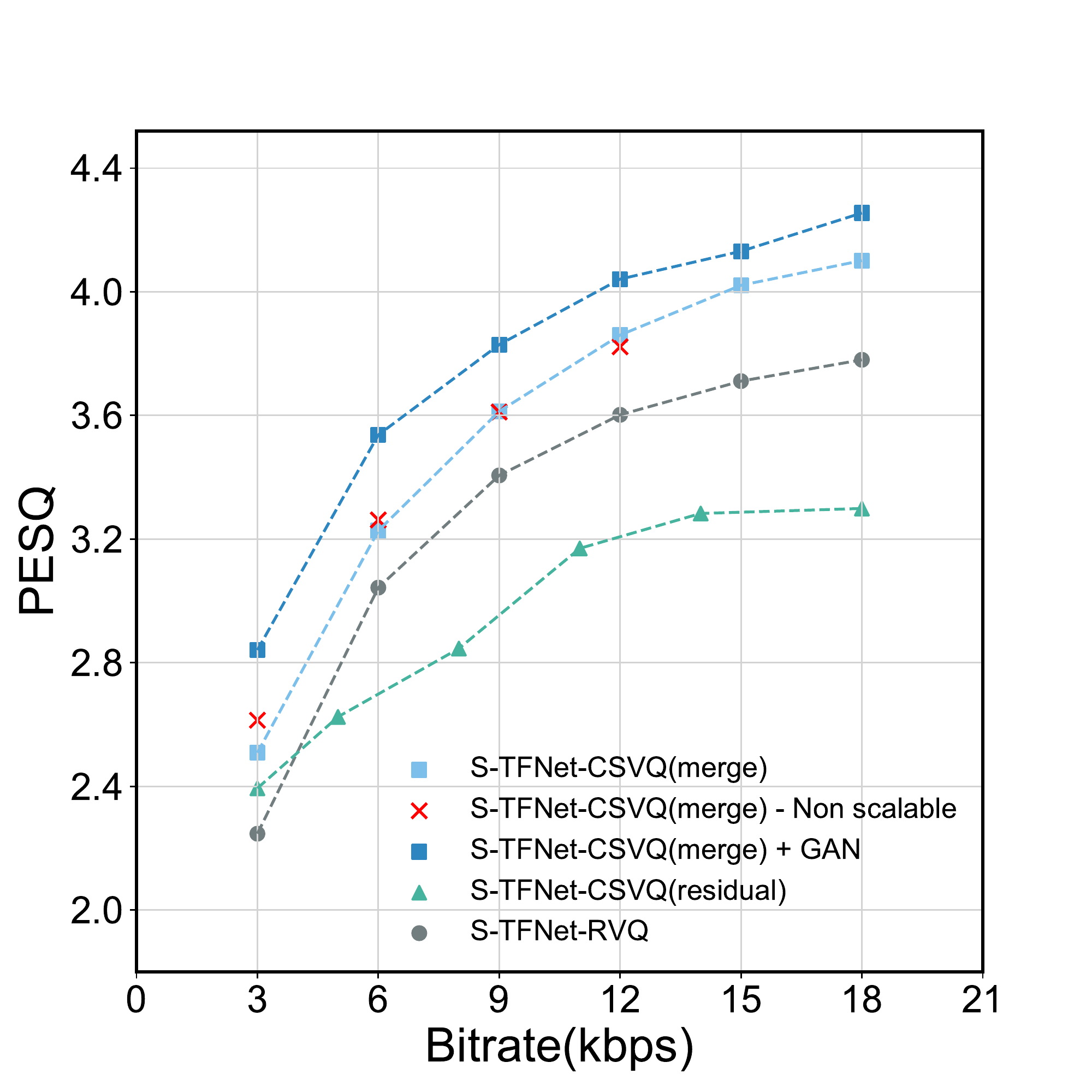}}
    \end{minipage}
\end{minipage}
\vspace{-0.4cm}
\caption{Subjective and objective evaluation results. The red dotted line in (a)(b) represents the score of the reference.}
\vspace{-0.2cm}
\label{fig:evaluation}
\vspace{-0.3cm}
\end{figure*} 
\vspace{-0.2cm}
\subsection{Cross-Scale Scalable Vector Quantization}
\label{ssec:scalablevq}
Vector Quantizer discretizes the learned features from encoding with a set of trainable codebooks according to the target bitrate. To reduce the codebook size when rate increases, several types of vector quantizers could be used like residual and product. The residual quantizer with multiple stages could also achieve bitrate scalability. However, the quantized features are at a fixed scale, typically the output of the encoder (see Figure \ref{fig:SVQ}(a)). Motivated by the spatial scalability in traditional scalable video coding and the U-Net structure, we propose to split bits at different scales of features and propose the cross-scale scalable vector quantization. It leverages a fuse-VQ-refine paradigm for transmitting multiple bitstreams at different scales within a single network.

As shown in Figure \ref{fig:model structure}(a), there are $N$ quantizers. At the lowest bitrate $B_0$ given by $Q_0$, only the encoder output is quantized and transmitted to reconstruct a coarse signal. With more bitstreams $B_1,...,B_i,i=1,2,...,N-1$, more details are recovered. To achieve that, the $i$-th ($i>0$) bitstream is generated by fusing the reconstructed feature from previous bitstreams at the decoder $\displaystyle \mathcal{F}_{dec}^{i}$ with corresponding feature at the encoder $\displaystyle \mathcal{F}_{enc}^{i}$ 
and quantizing by $Q_i$. The quantized feature $\displaystyle \mathcal{\hat{F}}_{merge}^{i}$ is then used to refine $\displaystyle \mathcal{F}_{dec}^{i}$ to generate an enhanced one $\displaystyle \mathcal{F}^{'i}_{dec}$. Specifically, we employ the fuse-VQ-refine module as shown in Figure \ref{fig:model structure}(b). Feature from the encoder $\displaystyle \mathcal{F}_{enc}^{i}$ is concatenated with that from the decoder $\displaystyle \mathcal{F}_{dec}^{i}$. They are then fed into two 2-D convolutional layers and downsampled by a 1-D convolutional layer to generate the merge feature $\displaystyle \mathcal{F}_{merge}^{i}$. During decoding, the quantized feature $\displaystyle  \mathcal{\hat F}_{merge}^{i}$ is first upsampled and then concatenated with $\displaystyle \mathcal{F}_{dec}^{i}$. The following two 2-D convolutional layers are trained to generate the refined feature $\displaystyle \mathcal{F}^{'i}_{dec}$. It should be noted that as partial decoding to get $\displaystyle \mathcal{F}_{dec}^{i}$ needs to be performed during encoding, when there are transmission errors, possible teacher forcing needs to be considered during training. We leave this to future work and don't consider transmission errors here.

As the resolution of feature maps changes in different layers of the encoder, in low-rate scenarios $\displaystyle \mathcal{F}_{merge}^{i}$ fused from deeper encoder layers contains high-level information of the audio and helps to produce a coarse reconstruction. In high-rate scenarios, $\displaystyle  \mathcal{F}_{merge}^{i}$ fused from early encoder layers with rich details helps to recover high frequency details.

We adopt a group quantization mechanism similar to \cite{baevski2019vq}, but without sharing codebooks across groups. Specifically, each transmitted feature is splitted into $G$ groups and each group is quantized with a separate codebook with $K$ codewords.  For the same $G$ and $K$ for all quantizers, each bitstream $B_i$ will consume a constant bit budget given by $R_i=G\log_2K$ bits. Taking $R=R_0+R_1$ with two bitstreams as an example, both $Q_0$ and $Q_1$ have six codebooks with 1024 codewords in each codebook. Four overlapped frames with 20ms new data are quantized together with the same codebook, so the consumed bitrate could be calculated as $2\times (6\times log_21024) /0.02 = 6$ kbps. Here no entropy coding is considered nor employed.   

Let $n$ denote the number of bitstreams needed to achieve the desired bitrate. During training, we randomly sample $n \in \{1,...,N\}$ with uniform distribution and only use the first $n$ quantizers \{$Q_0$,...,$Q_{n-1}$\} at each iteration for a minibatch. This enables a single model to operate at several target bitrates. We train the codebooks of each quantizer with exponential moving average, following the method proposed in \cite{oord2017neural}. During inference, when the $i$-th bitstream is transmitted, the $\displaystyle  \mathcal{\hat F}_{merge}^{i}$ is used to refine corresponding decoder feature; otherwise, $\displaystyle \mathcal{\hat F}_{merge}^{i}$ is set to zero for decoding.


\subsection{Training Objective}
\label{ssec:loss}
For good recovery quality, we use several objective terms during training, which is shown below
 \vspace{-0.1cm}
\begin{equation}
    \mathcal{L}=\mathcal{L}_{Comp}+\lambda_{1}\mathcal{L}_{Mel}+\lambda_{2} \mathcal{L}_{VQ}.
    \label{eq1}
\end{equation}
The first term $\mathcal{L}_{Comp}$ is the mean-square-error (MSE) loss on the power-law compressed STFT spectrum \cite{ephrat2018looking}. To keep STFT consistency \cite{wisdom2019differentiable}, the reconstructed spectrum is first transformed to time domain and then to the frequency domain to calculate the loss. The second term $\mathcal{L}_{Mel}$ is the multi-scale mel-spectrogram loss given by
\vspace{-0.1cm}
\begin{equation}
    \mathcal{L}_{Mel}=\mathbb{E}_s[\sum_{n=1}^W{||\phi^{n}(s)-\phi^{n}(\hat{s})||_{1}}],
    \label{eq2}
\end{equation}
where $\phi^{n}(\cdot)$ is the function that transforms a waveform into the mel-spectrogram using $n$-th window size. Following \cite{gritsenko2020spectral}, we calculate the mel spectra over a sequence of window-lengths between 64 and 2048. The last term $\mathcal{L}_{VQ}$ is the commitment loss similar to that in \cite{oord2017neural}, forcing the encoder to generate a representation approaching the selected codeword. The scalars $\lambda_{1}$ and $\lambda_{2}$  are weights to balance the three terms.

\section{Experiments}
\label{sec:experiment}
\subsection{Dataset and Settings}
\label{ssec:EXPconfig}
We take 150 hours of 16kHz multilingual clean speech from the DNS Challenge at ICASSP 2021 dataset \cite{DNSchallenge}. 
For training, we use 180000 clips, each of 3 seconds in duration. For evaluation, we use 1158 clips of 10s without any overlapping with the training data. Hanning window is used in STFT with a window length of 20 ms and a hop length of 5 ms. 

During training, we use adam optimizer with a learning rate of 0.0003. The network is trained for 200 epochs with a batch size of 80.

\vspace{-0.1cm}
\begin{figure*}[tb]
\vspace{-0.2cm}
\centering
\begin{minipage}[b]{0.2\linewidth}
  \centering
  \raisebox{-0.3\height}{\subfigure[Uncompressed signal]{\vspace{5cm}\includegraphics[width=1\linewidth]{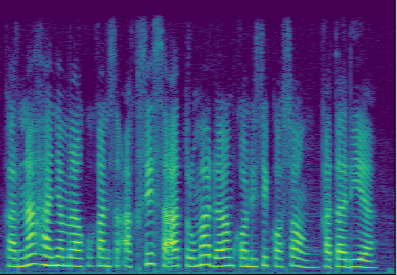}}}
\end{minipage}
\hspace{-0.5cm} 
\begin{minipage}{0.65\linewidth}
    \begin{minipage}[b]{1.0\linewidth}
      \centering
      \subfigcapskip=-2.5pt
      \subfigure[RVQ 3kbps]{\includegraphics[width=0.29\linewidth]{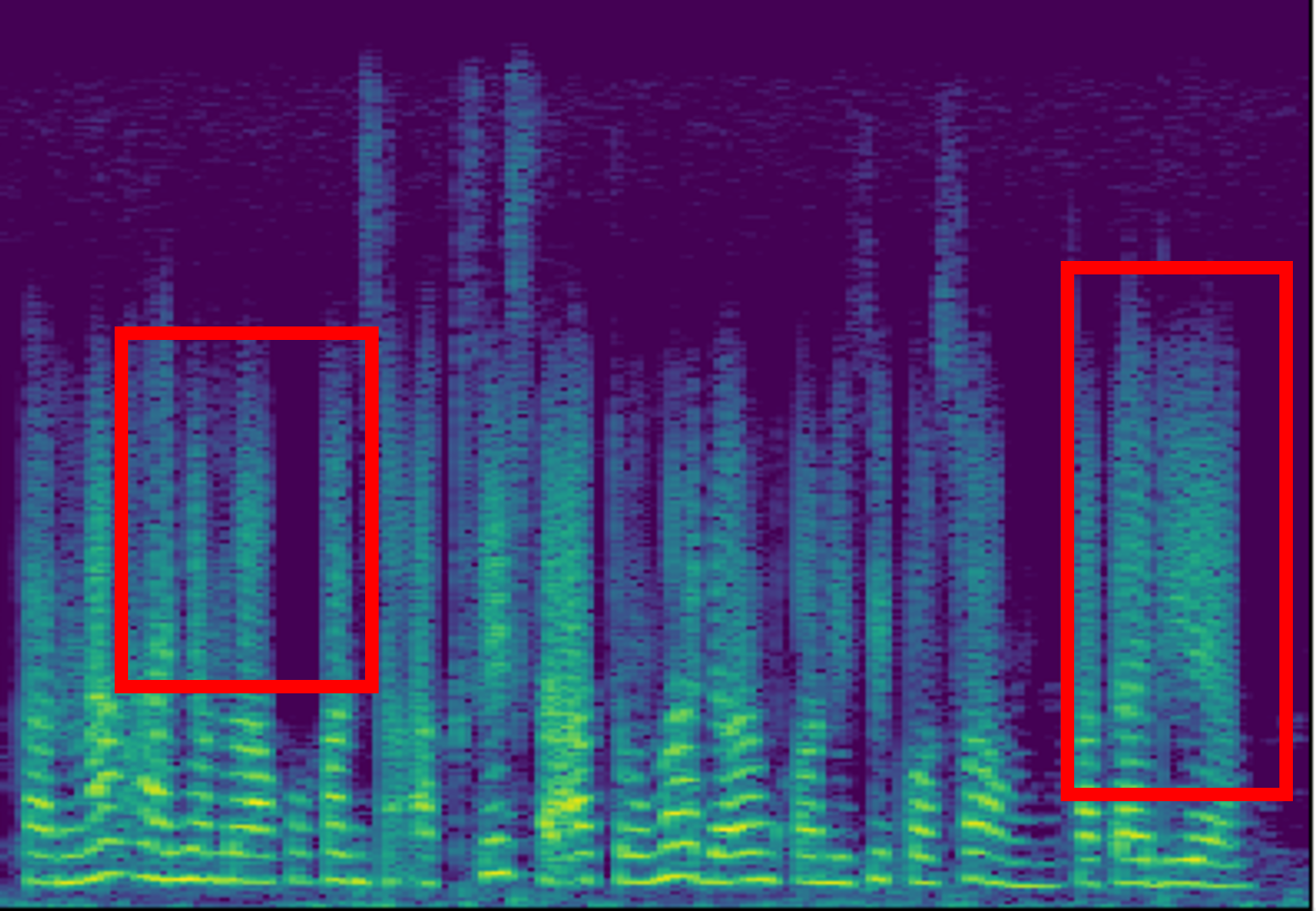}}
      \vspace{-0.1cm}
      \subfigure[RVQ 6kbps]{\includegraphics[width=0.29\linewidth]{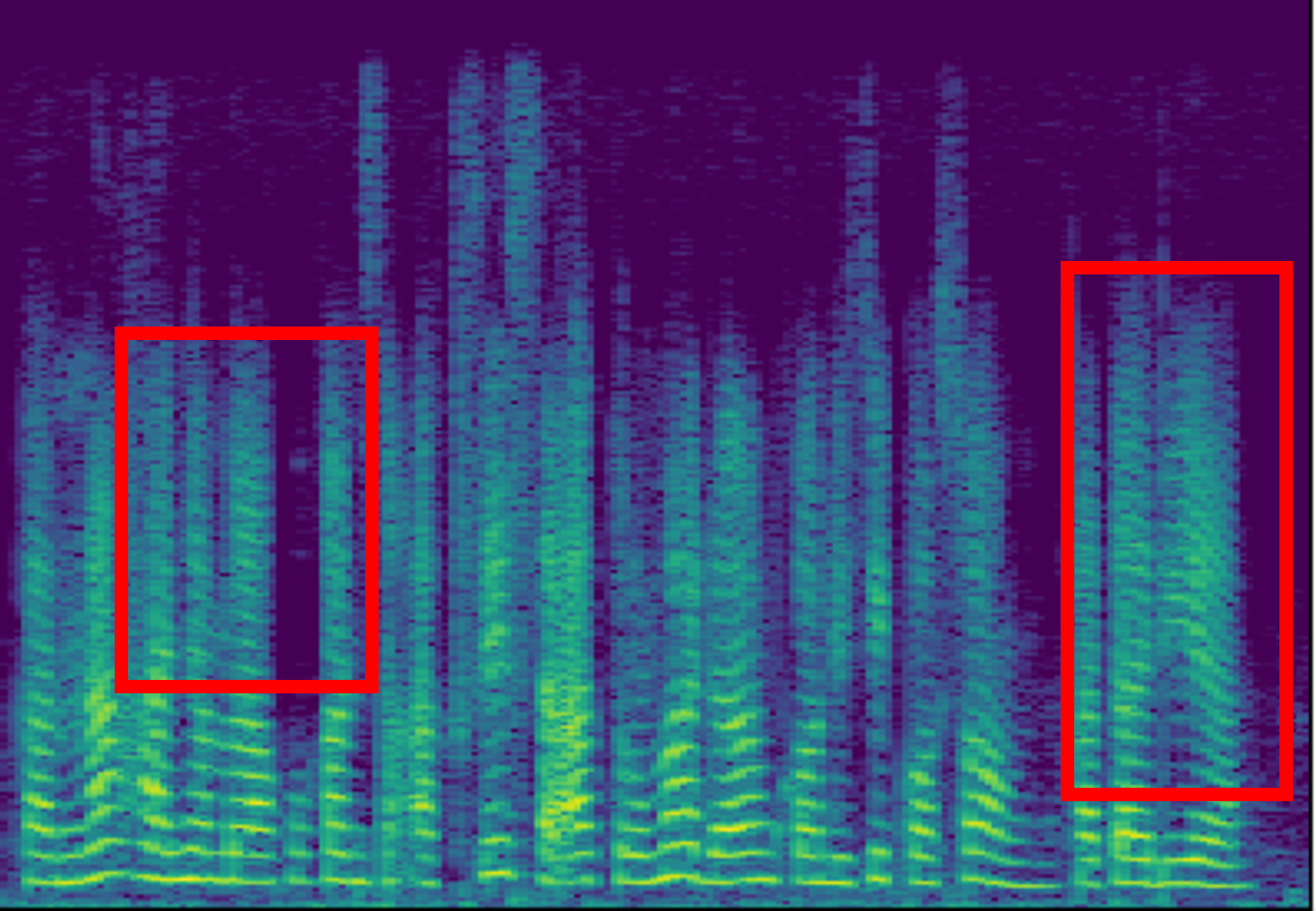}}
      \subfigure[RVQ 18kbps]{\includegraphics[width=0.29\linewidth]{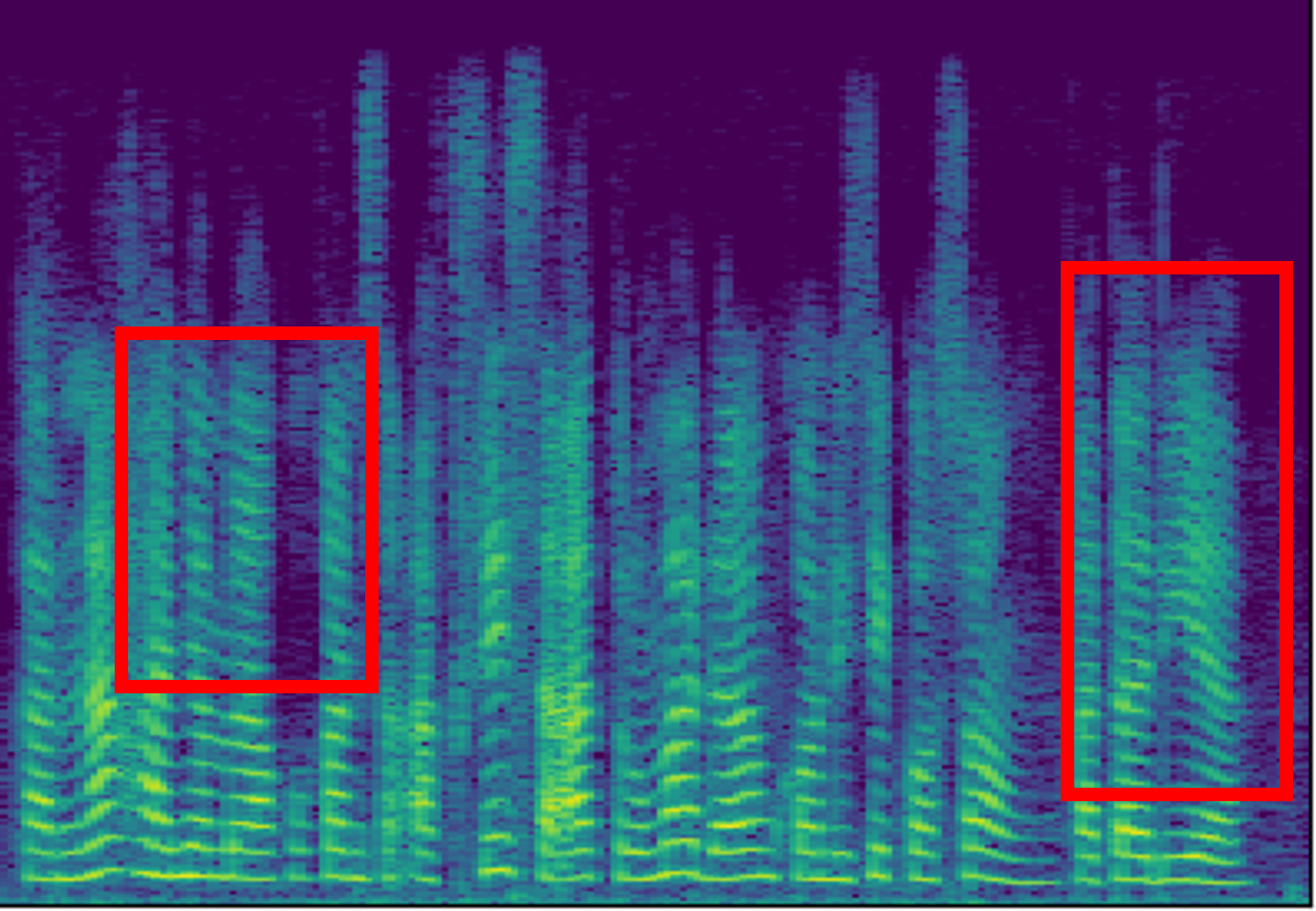}}
    \end{minipage}
    \begin{minipage}[b]{1.0\linewidth}
      \centering
      \vspace{-0.05cm}
      \subfigcapskip=-2.5pt
      \subfigure[CSVQ 3kbps]{\includegraphics[width=0.29\linewidth]{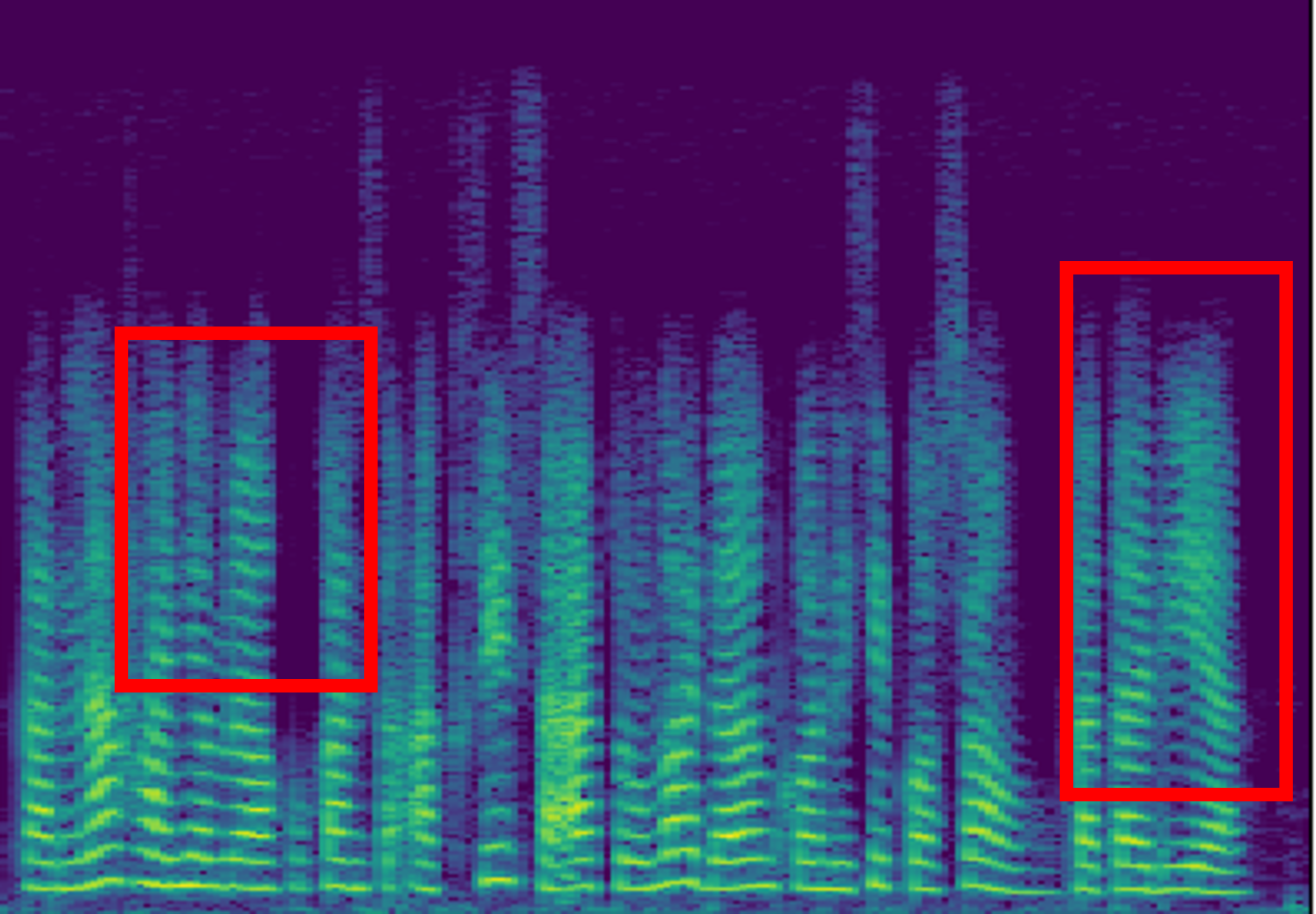}}
      \subfigure[CSVQ 6kbps]{\includegraphics[width=0.29\linewidth]{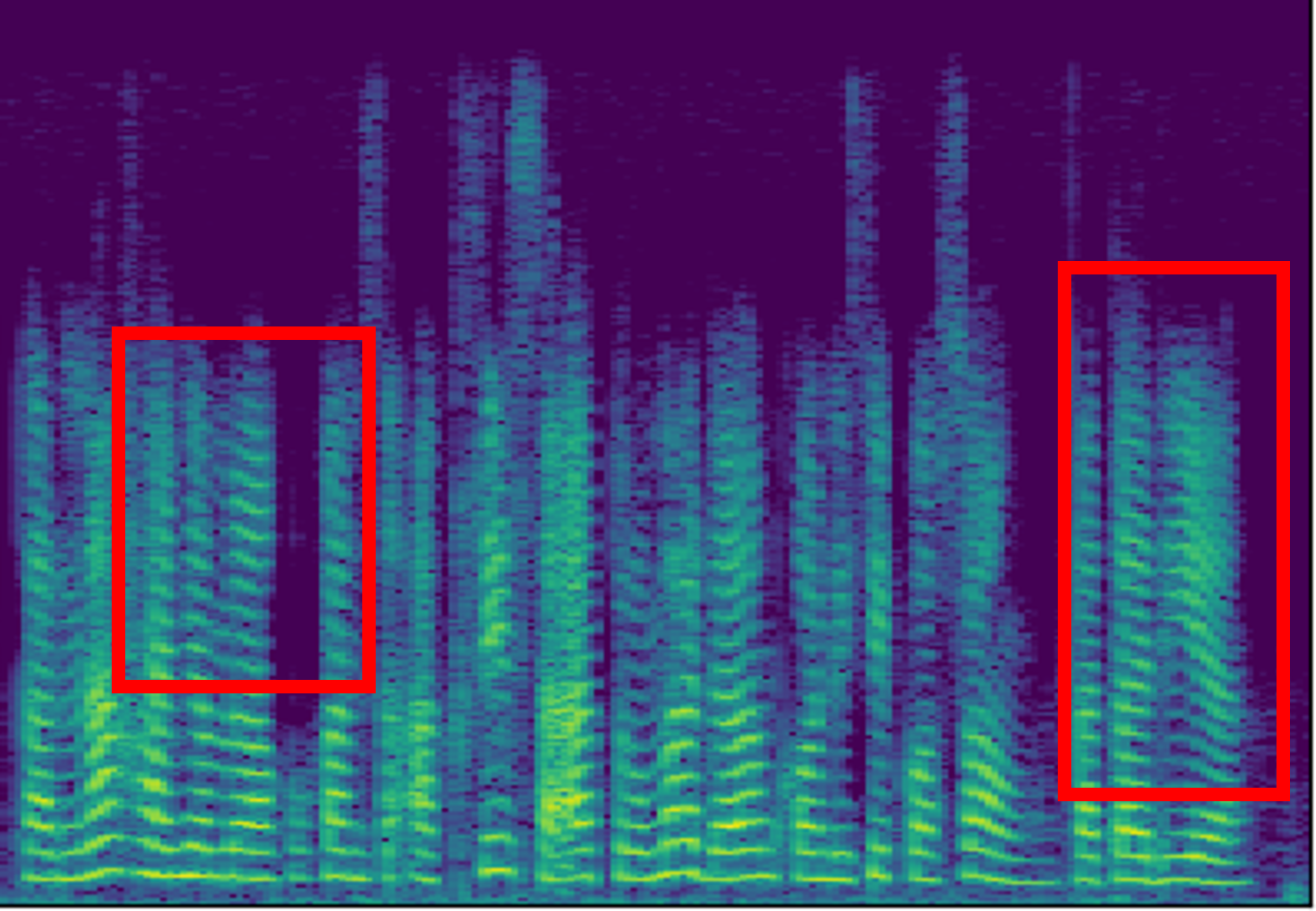}}
      \subfigure[CSVQ 18kbps]{\includegraphics[width=0.29\linewidth]{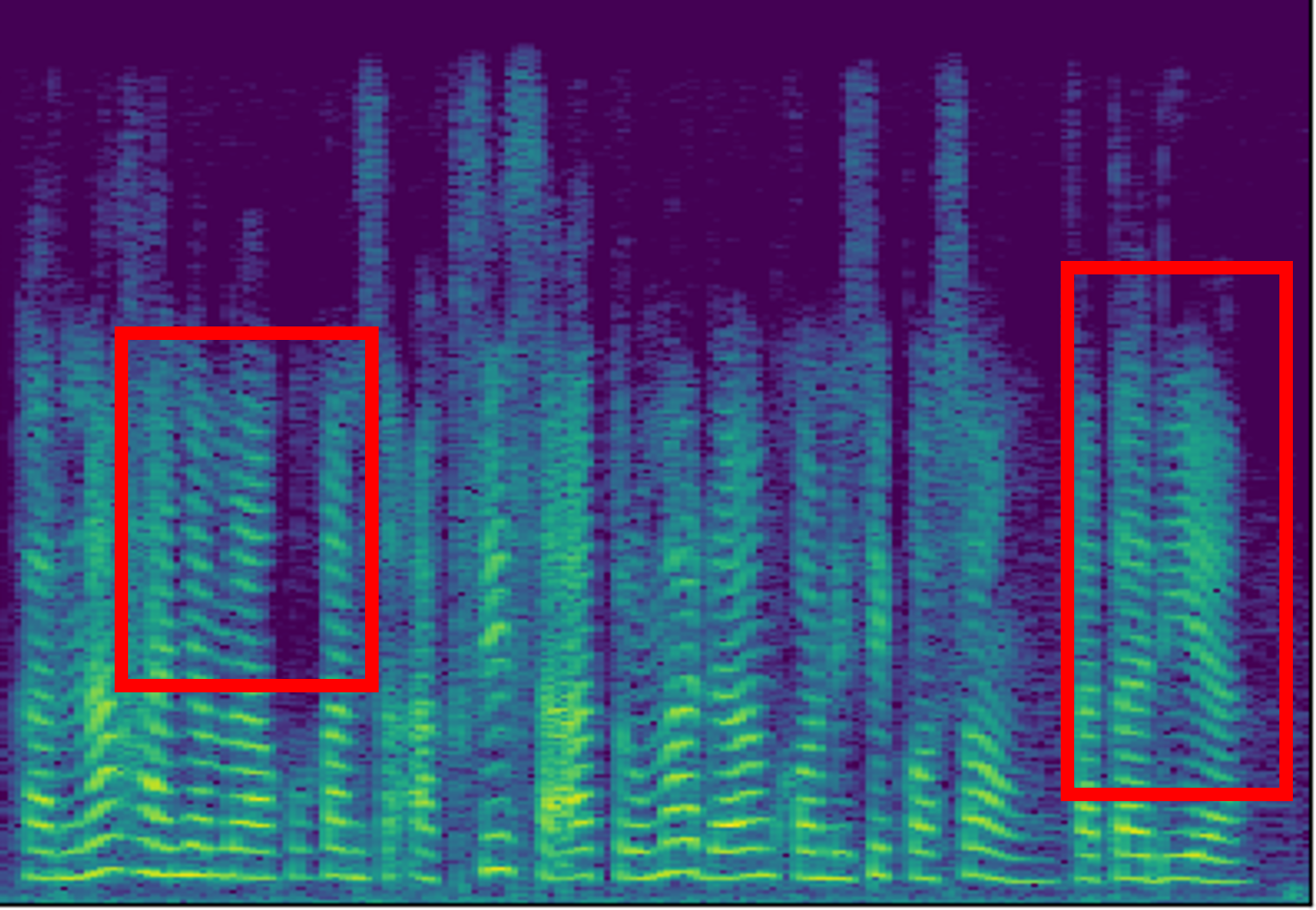}}
    \end{minipage}
\end{minipage}
\vspace{-0.4cm}
\caption{Visualization.}
\vspace{-0.5cm}
\label{fig:Visual}
\vspace{-0.2cm}
\end{figure*} 

\subsection{Subjective Quality Evaluation}
\label{ssec:EXPsubjective}
To evaluate the reconstructed signal, we conduct a subjective listening test with a MUSHRA-inspired crowd-sourced methodology \cite{subjective}, where 8 participants evaluate 30 samples. In MUSHRA evaluations, the listener is presented with a hidden reference and a set of test samples by different codecs. The anchor based on low-pass filter is not used in our experiment. 

Figure \ref{fig:evaluation}(a)(b) shows the subjective evaluation results, where we compare our bitrate-scalable method with two real-time audio codecs, Opus and Lyra. Opus \cite{valin2012definition} is a versatile codec that is widely used for real-time communications, supporting narrowband to fullband speech and audio with bitrate from 6kbps to 510kbps. Lyra \cite{kleijn2021generative} is an autoregressive generative speech codec proposed recently, which reconstructs high quality speech at 3 kbps.


\textbf{Low bitrate scenarios} Figure \ref{fig:evaluation}(a) shows the quality versus bitrate evaluation at low bitrate range. It can be seen that our bitrate-scalable model with the proposed CVSQ at 3 kbps (denoted by CSVQ3k) significantly outperforms both Opus at 6 kbps and Lyra at 3kbps, and also performs better than Opus at 9kbps. To better verify the efficiency of the proposed CSVQ scheme, we also implement a RVQ-based scalable method based on the same backbone as CSVQ. It is shown that CSVQ outperforms RVQ at 6kbps, indicating that the rich information carried by multi-scale features indeed helps the reconstruction quality. Although this paper focuses only on the CSVQ technique, 
to show its potential when combined with other sophisticated techniques in the literature, we also implement a CSVQ scheme with adversarial training, shown as CSVQ+GAN in Figure \ref{fig:evaluation}(a). At 3kbps, the CSVQ+GAN largely outperforms CSVQ and it also exceeds Opus at 9kbps by a large margin.

\textbf{High bitrate scenarios} Figure \ref{fig:evaluation}(b) shows the comparison at high bitrates. It is shown that CSVQ at 18kps performs slightly better than Opus at 20kbps, indicating that our model could achieve a high quality close to transparent at 18kbps.

\vspace{-0.1cm}
\subsection{Ablation Study}
For ablation study, we use PESQ \cite{pesq} for quality evaluation. Although it is not proposed for coding quality evalution, we found that when using the same backbone and loss function, it matches our perceptual quality well. Figure \ref{fig:evaluation}(c) shows the quality versus bitrate of various schemes. 
\label{ssec:EXPobjective}
\subsubsection{Bitrate scalability}
\label{sssec:subsubhead}
We evaluate the CSVQ with the bitrate $R$ ranging from 3kbps to 18 kbps. We investigate two different configurations: a) a scalable solution trained with random bitrate sampling and evaluated at bitrate $R$ (denoted as S-TFNet-CSVQ (merge) in Figure \ref{fig:evaluation}(c)); b) a non-scalable model trained and evaluated at a fixed bitrate $R$ (denoted as S-TFNet-CSVQ (merge)-Non scalable). 

It can be seen that the S-TFNet with proposed CSVQ provides a good tradeoff between quality and bitrate and achieves an obvious quality improvement with the increasing bitrate. Furthermore, 
the bitrate-scalable model is on par with the non-scalable one, showing the effectiveness of random bitrate sampling during training.

\subsubsection{CSVQ vs. RVQ}
\label{sssec:subsubhead}
We further compare the CSVQ with the common RVQ used in \cite{zeghidour2021soundstream} across a wide bitrate range from 3 kbps to 18 kbps. It is shown that when using the same backbone, the proposed CSVQ consistently outperforms RVQ at different bitrates.

Figure \ref{fig:Visual} illustrates the spectrogram of the signal reconstructed by CSVQ and RVQ at different birates. It can be seen that CSVQ produces an output with clearer harmonics than RVQ when both operating at the same bitrate. Additionally, we can observe that more high-frequency details are recovered with the increasing bitrate, indicating that the additional information by $B_1,B_2,...,B_{N-1}$ carrying rich details can progressively refine the output quality.

\subsubsection{Ablation study on fuse-VQ-refine module }
\label{sssec:ablation}
Inspired by RVQ, we also investigate a residual variant of the fuse-VQ-refine module (see Figure \ref{fig:model structure}(c)), denoted as S-TFNet-CSVQ(residual) in Figure \ref{fig:evaluation}(c). $Q_1$,...,$Q_{N-1}$ quantize the feature residual between the encoder and decoder instead of learned fused information. To make the residual sparse, an $\ell_2$ feature loss is introduced to enforce the decoder feature $\displaystyle \mathcal{F}_{dec}^{i}$ be close to the encoder feature $\displaystyle \mathcal{F}_{enc}^{i}$ of the corresponding layer.

From Figure \ref{fig:evaluation}(c) we can see that the residual variant shows a similar trend as the merge scheme S-TFNet-CSVQ(merge) that the quality increases with the bitrate. However, the residual variant is much poorer than the proposed merge scheme for CSVQ and this gap gets larger as the bitrate increases. This demonstrates that the merge scheme automatically learns more meaningful information from the multi-scale concatenated features of the encoder and decoder, which could better refine the decoder features than only using the residual information. 
\vspace{-0.1cm}
\subsubsection{Combination with adversarial training}
In Figure \ref{fig:evaluation}(a) we show the quality boost when combining CSVQ with GAN for training at 3kbps. Here we show the comparison across the wide bitrate range. For discriminators, we use STFT spectrum as the input and four 2D convolutional layers with reduced resolutions in both time and frequency dimensions for extracting features, followed by a fully-connected layer and temporal pooling to produce logits. We use the least-square loss as the GAN objective (LSGAN\cite{LSGAN}). Figure \ref{fig:evaluation}(c) shows that the adversarial training consistently boosts the quality at various bitrates, indicating the potential of the proposed CSVQ to be combined with other sophisticated techniques for audio/speech coding.

\section{Conclusions}
\label{sec:Conclusion}
We propose the CSVQ, a cross-scale scalable vector quantization that can achieve bitrate scalability with good rate-distortion performance. Both subjective and objective experiments demonstrate its coding efficiency and great potential to be combined with other sophisticated techniques. Although we use speech coding based on TFNet backbone as an example in this paper, it could be applied to general audio coding and any other mirror-like auto-encoder based neural audio coding as well.

\bibliographystyle{IEEEtran}
\bibliography{mybib}

\end{document}